\documentclass[journal]{IEEEtran}

\usepackage[numbers,sort&compress]{natbib}
\usepackage{amsmath}
\usepackage{stmaryrd}
\usepackage{amssymb}
\usepackage{graphicx}
\usepackage{epstopdf}
\usepackage{stfloats}
\usepackage{dsfont}

\graphicspath{{fig/}}
\DeclareGraphicsExtensions{.eps}
\usepackage{epstopdf}
\usepackage{algorithm}
\usepackage{algorithmic}

\begin{document}
%
\title{Polar Coded Modulation with Optimal Constellation Labeling}
%
%
%
\author{Kai~Chen,
        Kai~Niu,
        and~Jia-Ru~Lin, 

\thanks{The authors are with the Key Laboratory of Universal Wireless Communication, Ministry of Education,
Beijing University of Posts and Telecommunications,
Beijing 100876, China. (e-mail: kaichen@bupt.edu.cn)}
}

%

\maketitle

\newtheorem{theorem}{Theorem}
\newtheorem{example}{Example}

\begin{abstract}

A practical $2^m$-ary polar coded modulation (PCM) scheme with optimal constellation labeling is proposed.
To efficiently find the optimal labeling rule, the search space is reduced by exploiting the symmetry properties of the channels.
Simulation results show that the proposed PCM scheme can outperform the bit-interleaved turbo coded modulation scheme used in the WCDMA (Wideband Code Division Multiple Access) mobile communication systems by up to 1.5dB.

\end{abstract}


\section{Introduction}
\IEEEPARstart{P}{olar} codes \cite{pcode}, which have been proven to achieve the symmetric capacity of binary-input discrete memoryless channels under the successive cancellation (SC) decoding, is of great interest recently.
A bit-interleaved polar coded modulation (BIPCM) scheme is discussed in \cite{pocm} to improve the spectral efficiency.
Inspired by \cite{mlvl}, a multi-level polar coding framework is described by Seidl et al. \cite{mlvlpc} by viewing the dependencies between the bits mapped to a single modulation symbol as a kind of channel transformation.
In this letter, we optimize the polar coded modulation (PCM) scheme under the framework of multi-level coding by searching the optimal constellation labeling. The search space is reduced by exploiting the symmetry properties of the polarized channels.
Simulation results of 8-ary pulse amplitude modulation (PAM) under SC decoding over additive white Gaussian noise (AWGN) channels show that, the PCM scheme with optimal labeling provides an improvement of about $1.5$dB over the BIPCM scheme.
By utilizing the improved SC decoding techniques \cite{scl1}, \cite{scl}, \cite{cadec}, the proposed PCM scheme can be further improved and outperforms the bit-interleaved turbo coded modulation (BITCM) scheme used in WCDMA system \cite{3gpp} by up to 1.5dB.

\section{Polar Coded Modulation}
\label{section_RCPrP}
In this paper, we use calligraphic characters, such as $\mathcal{X}$ and $\mathcal{Y}$, to denote sets.
We write the Cartesian product of $\mathcal{X}$ and $\mathcal{Y}$ as $\mathcal{X} \times \mathcal{Y}$ and the $n$-th Cartesian power of $\mathcal{X}$ as ${\mathcal{X}}^n$.
We use notation $x_1^N$ to denote an $N$-dimensional vector $\left(x_1, x_2, \cdots, x_N \right)$ and $x_i^j$ to denote a subvector $\left(x_i, x_{i+1},\cdots, x_{j-1}, x_j\right)$ of $x_1^N$, $1\leq i \leq j \leq N$.

We write \mbox{$W: \mathcal{X} \mapsto \mathcal{Y}$} to denote a $2^m$-ary input memoryless channel with input alphabet $\mathcal{X}$, output alphabet $\mathcal{Y}$, and transition probabilities $W(y|x)$, $x \in \mathcal{X}$, $y \in \mathcal{Y}$.
The number of elements in $\mathcal{X}$ is $|\mathcal{X}|=2^m$.
When the channel $W$ is a binary-input memoryless channel (BMC), i.e., $m=1$, $\mathcal{X}=\{0,1\}$.
In $2^m$-ary modulation system, every $m$ bits form a vector ${{b}}^{m}_{1} \in \{0,1\}^m$, and are modulated into a modulation symbol $x \in \mathcal{X}$ under a one-to-one mapping called \emph{constellation labeling}
\begin{equation}
\label{equ_labeling}
L : \{0,1\}^{m} \mapsto \mathcal{X}
\end{equation}
Under a certain labeling rule $L$, we abuse notation slightly and use \mbox{$W: \{0,1\}^m \mapsto \mathcal{Y}$} to denote the channel with transition probabilities
\begin{equation}
W(y|{b}_{1}^{m})=W(y|x)
\end{equation}

\begin{figure}[!t]
\centering{
\includegraphics[width=90mm]{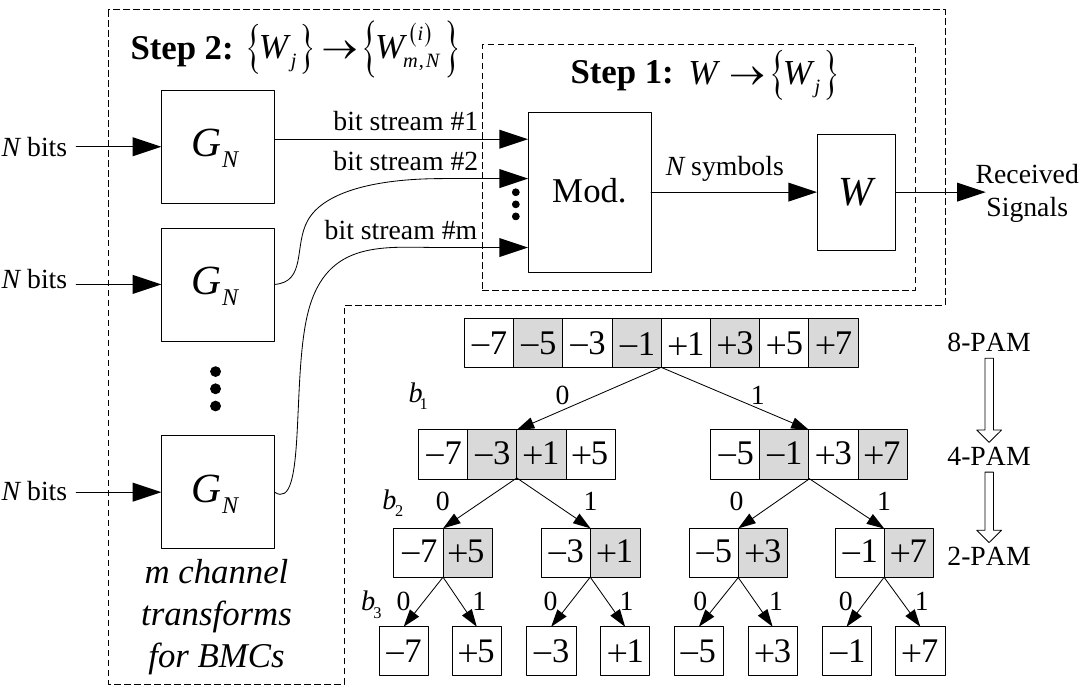}}
\caption{A diagram of the channel transform under PCM and an example of recursive constellation labeling over 8-PAM.}
\label{fig_example}
\end{figure}

A PCM transmission scheme under the multi-level coding framework is constructed by a two-step channel transform.
The first step is to transform the $2^m$-ary input channel $W$ into $m$ synthesized BMCs $W_j: \{0, 1\} \mapsto \mathcal{Y} \times \{0, 1\}^{j-1}$, where $j=1, 2, \cdots, m$.
The transition probabilities of $W_j$ are
\begin{equation}
\label{equ_wj}
W_j\left( y, {b}_{1}^{j-1}|{{b}_{j}} \right)=\sum\limits_{{b}_{j+1}^{m}\in {{\left\{ 0,1 \right\}}^{m-j}}}{\frac{1}{{{2}^{m-1}}}\cdot W\left( y|b_{1}^{m} \right)}
\end{equation}
The second step is performing a conventional $N$-dimensional channel polarization transform $G_N$ \cite{pcode} on each of these $m$ BMCs $W_j$, where $N=2^n$, $n=1, 2, \cdots$.
The resulting BMCs are denoted by $W_{m, N}^{(i)}: \{0, 1\} \mapsto \mathcal{Y}^N \times \{0, 1\}^{i-1}$, where $i=1, 2, \cdots, mN$.
The BMC $W_{m, N}^{(i)}$ with superscript $i$ is derived from $W_j$ with $j=\lceil i/N \rceil$, where ceiling function $\lceil x \rceil$ is the smallest integer not less than $x$.
After calculating the reliabilities of all the $m N$ BMCs, the $K$ most reliable ones are selected to carry the information bits, and others are fixed to frozen bits which is known both to the encoder and the decoder.
The second step of the channel transform can be viewed as constructing a set of \mbox{$N$-length} polar codes on each of the $m$ BMCs in set $\{W_j\}$ with a total rate $R=\frac{K}{mN}$.
A diagram of the channel transform under the PCM scheme is shown in \mbox{Fig. \ref{fig_example}}.

\section{Optimal Constellation Labeling}
\label{section_harq}
Given a $2^m$-ary input memoryless channel $W$ and a labeling rule $L$, let $I(W)$ denote the symmetric capacity, i.e., the mutual information between the input and output of $W$ when the channel input letters are used with equal probability.
The symmetric capacities of $W_j$ can be calculated as
\begin{eqnarray}
\label{equ_IWj}
I(W_j)&=&\sum\limits_{b_{1}^{m}\in {{\left\{ 0,1 \right\}}^{m}}}{\sum\limits_{y\in \mathcal{Y}}{\left( \text{Pr}\left( y,b_{1}^{m} \right)\cdot \log \frac{\text{Pr}\left( y|b_{1}^{j} \right)}{\text{Pr}\left( y|b_{1}^{j-1} \right)} \right)}} \nonumber \\
&&
\end{eqnarray}
After adding up the symmetric capacities $I(W_j)$ for all $j \in \{1,2,\cdots,m\}$, a total capacity is obtained
\begin{eqnarray}
\sum\limits_{j=1}^{m}{I\left( {{W}_{j}} \right)}&=&\sum\limits_{b_{1}^{m}\in {{\left\{ 0,1 \right\}}^{m}}}{\sum\limits_{y\in \mathcal{Y}}{\left( \text{Pr}\left( y,b_{1}^{m} \right)\cdot \log \frac{W\left( y|b_{1}^{m} \right)}{\text{Pr}\left( y \right)} \right)}}\nonumber \\
\label{equ_sum}
&=&I\left( W \right)
\end{eqnarray}

Equation (\ref{equ_sum}) is in fact the chain rule of mutual information which reveals the following two facts:

1. The channel transform from the $2^m$-ary channel $W$ to the $m$ synthesized BMCs $\{W_j\}$ will not bring any (symmetric) capacity loss;

2. The total capacity is always $I(W)$ and not affected by the selection of labeling rules.
\medskip

\begin{figure}[!b]
\centering{
\includegraphics[width=80mm]{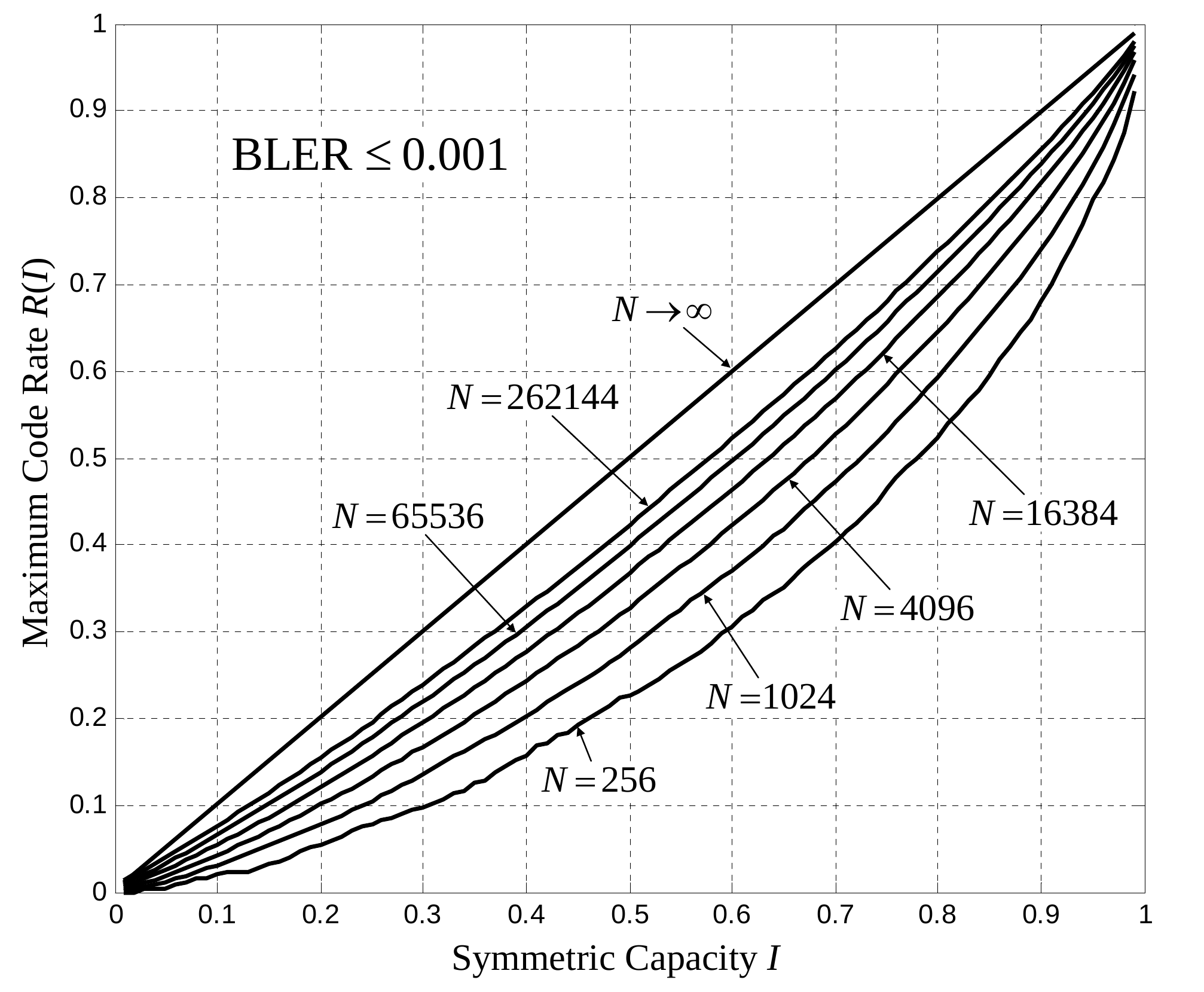}}
\caption{Maximum code rates of polar codes over binary-input AWGN channel when the block error probability is less than $0.001$.}
\label{fig_IWvsR}
\end{figure}

However, for the finite-length polar coding, the practical code rate cannot achieve the symmetric capacity.
Fig. \ref{fig_IWvsR} shows the maximum code rate $R(I)$ of polar code over binary-input AWGN channel when the symmetric capacity $I$ takes value in $(0,1)$ and the block error rate (BLER) is less than $0.001$.
When the code length takes a practical value and is finite, the gap between the capacity and the maximum achievable code rate $I-R(I)$ differs as $I$ changes.
According to (\ref{equ_IWj}), although the total capacity is fixed, the distribution of $\{I(W_j)\}$ can be adjusted by changing the labeling rule.
Therefore, the performance of a practical PCM transmission system can be improved by optimizing the constellation labeling rule.

Given a specific labeling rule, the BLER performance of a $2^m$-ary PCM scheme over AWGN channel can be evaluated using density evolution \cite{de}.
Obviously, a brute-force search for the optimal labeling rule needs $(2^m)!$ tentative constructions of PCM.

The labeling problem can be described in a recursive manner as follows.
For a $2^m$-ary labeling problem with $m \ge 2$, half of the $2^m$ modulation symbols are selected from $\mathcal{X}$ and these symbols are corresponding to vectors $b_1^m \in \{0,1\}^m$ with $b_m=0$, and the others with $b_m=1$.
The number of possible selections is $\left(\begin{matrix}
   {{2}^{m}}  \\
   {{2}^{m-1}}  \\
\end{matrix}\right)$.
Then, the problem is degraded to labeling two \mbox{$2^{(m-1)}$-ary} component constellations.
The component constellations are usually non-uniform and with a DC (direct current) bias.
Each of the two sub-problems are further decomposed into two problems of $2^{(m-2)}$-ary labeling.
This procedure continues until the number of sub-problems increases to $2^{m}$, each of the component constellations has only one symbol.
Fig. \ref{fig_example} shows an example of an 8-PAM labeling.

Note that the symmetric capacity of a binary-input channel $W_j$ will not be changed by swapping the labeling rule of the input letters $b_j=0$ and $b_j=1$ (see (\ref{equ_IWj})).
Thus, the search space of each component constellation can be halved.
Therefore, the number of candidate labeling rules for a $2^m$-ary PCM system is
\begin{eqnarray}
\label{equ_sspc}
S\left( m \right)=\prod\limits_{i=1}^{m}{{{\left( \frac{1}{2}\left(
    \begin{matrix}
   {{2}^{i}}  \\
   {{2}^{i-1}}  \\
\end{matrix} \right) \right)}^{{{2}^{m-i}}}}}
   =\frac{(2^m)!}{2^{(2^m-1)}}
\end{eqnarray}

In this way, the number of trials to find the optimal labeling rule is reduced by a factor of $2^{(2^m-1)}$.
For example, the search space for the optimal 8-ary PCM labeling is reduced from $8!=40320$ to $S(3)=315$.

\subsection{Simulation Results}
\label{section_harq:design}
\begin{figure}[!t]
\centering{
\includegraphics[width=80mm]{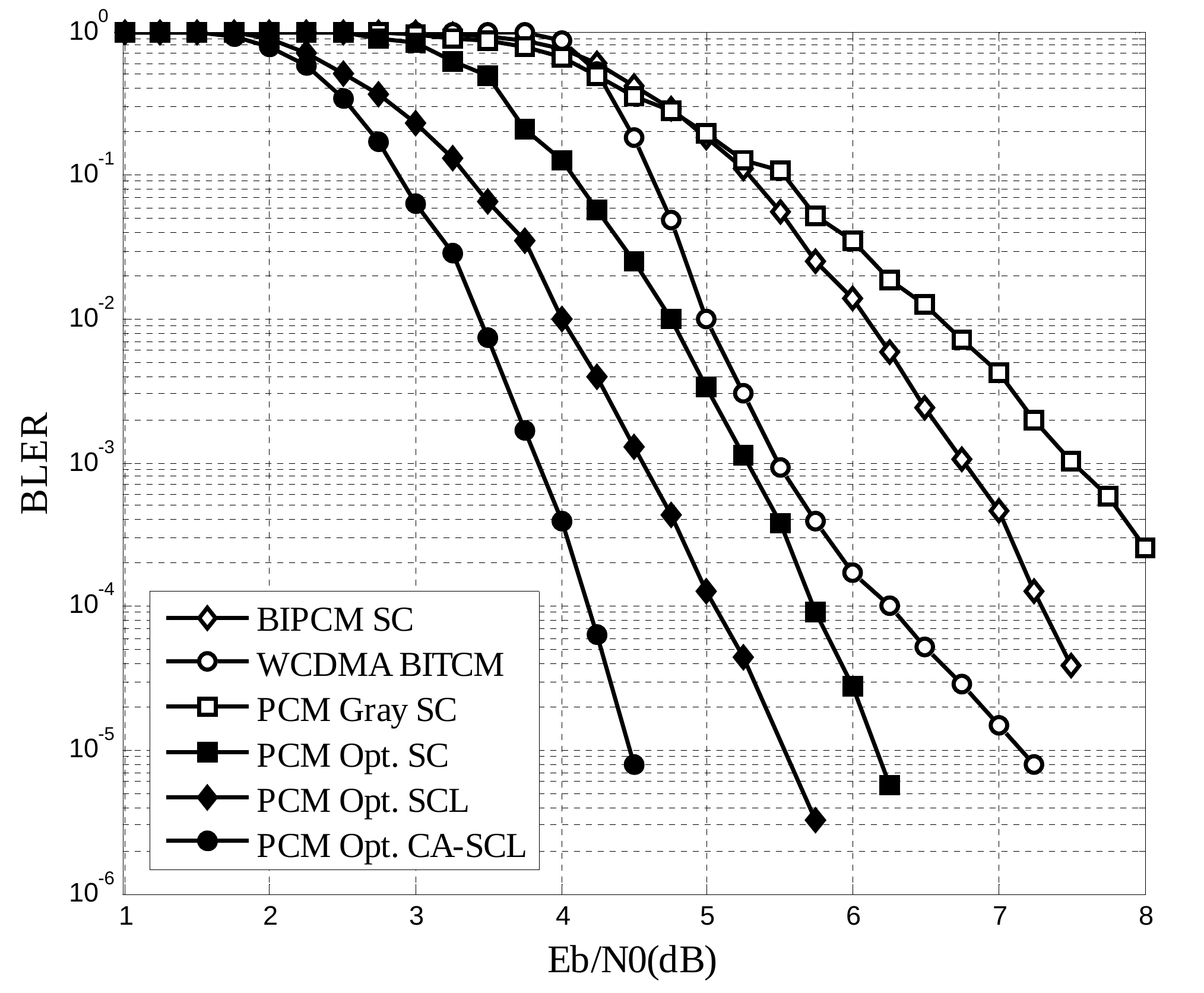}}
\caption{BLER performance of the PCM schemes of 8-PAM over AWGN channels with transmission rate $1$ bit/dim.}
\label{fig_perform}
\end{figure}

Fig.\ref{fig_perform} gives the BLER performance over 8-PAM with optimal labeling of the proposed PCM.
The information blocks consist of $K=512$ bits and the number of transmitted symbols is set at $N=512$.
The code length is defined as the number of the bits which are mapped to symbols, i.e., $mN=1536$.
Thus, code rate is $\frac{K}{mN}=\frac{1}{3}$ and the transmission rate is $\frac{K}{N}=1$ bit/dim.
The SC decoding algorithm and its two improved decoding schemes, namely, the successive cancellation list (SCL) decoding \cite{scl1}, \cite{scl}, and the CRC(cyclic redundant check)-aided SCL (CA-SCL) decoding are used to decode the polar codes, where the search width is set at $32$.
In comparison, the performance of the BIPCM scheme \cite{pocm}, the BITCM scheme used in the WCDMA systems \cite{3gpp}, and the PCM scheme with Gray labeling are also provided.
The simulation results show that the performance of PCM scheme relies significantly on the labeling rules.
Under the SC decoding, the optimal labeled PCM provides about 1.5dB improvement over the BIPCM scheme.
Under the CA-SCL decoding, the proposed PCM scheme can be further improved and outperforms the BITCM scheme by up to $1.5$dB.
The figure also shows that the PCM has no sign of error floor down to the BLERs of $10^{-4}$.

\section{Conclusions}
\label{section_conclusions}
In this letter, a $2^m$-ary polar coded modulation (PCM) scheme under the framework of multi-level coding with the optimal constellation labeling is proposed.
When the code length is finite, the performance of PCM scheme can be significantly improved by optimizing the labeling rule.
Simulation results over AWGN channels show that, the proposed PCM scheme with improved decoding techniques can outperform turbo coded modulation scheme used in WCDMA system by up to 1.5dB.


%

\ifCLASSOPTIONcaptionsoff
  \newpage
\fi

\end{document}